\documentclass[aps,prl,twocolumn]{revtex4}
\usepackage{amssymb}
\usepackage[utf8]{inputenc}
\usepackage{epsf}

\begin{document}
\title{Electric-field-dependent energy structure of quasi-one-dimensional conductor {\it o}-TaS$_3$}
\author{V.F.~Nasretdinova, S.V.~Zaitsev-Zotov}
\affiliation{Kotel'nikov Institute of Radioengineering and Electronics of RAS, 125009 
Mokhovaya 11, Moscow, Russia}
\date{\today}

\begin{abstract}
Energy structure of the Peierls gap in orthorhombic TaS$_3$ is examined by spectral study of photoconduction. The gap edge and energy levels inside the Peierls gap are observed. The amplitude of the energy levels is found to depend on both the temperature and the electric field. The electric field of the order of 10 V/cm affects the energy levels and leads to the redistribution of intensity between peaks. The small value of the electric field indicates participation of the collective state in formation of the energy levels inside the Peierls gap.
\end{abstract}


\maketitle 

\section{Introduction}

Instability of the electron-phonon system of quasi-one-dimensional (quasi-1D) metal with respect to the crystal lattice distortion with the wave vector $Q=2k_F$, where $k_F$ is the Fermi momentum of electrons, leads to formation of a charge density wave (CDW) and to opening of the Peierls gap at the Peierls transition temperature $T_P$ \cite{cdwreview}. This single-particle gap becomes apparent from activation temperature-dependent conduction and Hall effect at $T<T_P$, from tunneling experiments, optical measurements, {\it etc.}

Impurities lead to pinning of the CDW and, therefore, play the central role in nonlinear collective transport in quasi-1D conductors \cite{cdwreview}. In addition, they affect the electronic states.  Effect of impurities on the electronic states was considered by T\"{u}tt\"{o} and Zawadovski \cite{zavadowski}. It was shown, that a charged impurity produces a couple of electronic levels near the gap edges. Position of this levels depends on the backscattering amplitude and the CDW phase on the impurity site. To our knowledge, no experimental evidence for impurity levels in quasi-1D conductors is found up to now.

Orthorhombic TaS$_3$ (o-TaS$_3$) can be considered as a model quasi-1D conductor. It has $T_P=220$~K. As for the Peierls gap value, the available data scatter. The data on the activation energy at a temperature range $100{\rm\ K}<T<T_P$ vary from 600~K ($2\Delta=0.1$~eV) \cite{sambongi} up to 800 -- 850~K ($2\Delta=0.14$~eV) \cite{cdwreview} for conduction measurements, and is 1000~K ($2\Delta=0.17$~eV) for the Hall effect \cite{TaS3hall}. Bolometric response studies \cite{naditkis,brill} give temperature-dependent value of gap defined as the onset of absorption, $2\Delta=0.125-0.127$~eV at $T \approx 100$~K and $2\Delta=0.15$~eV at 20~K \cite{naditkis,brill}. Interchain tunneling measurements provide a bigger gap value at temperatures $100{\rm\ K}<T<T_P$ defined as the maximum of tunnel conduction $2\Delta=0.18$~eV while the onset of tunneling is at $0.2\Delta=0.036$~eV\cite{latnewgap}. The activation energy for low-temperature single-particle conduction obtained from the temperature-dependent photoconduction is 1250~K ($2\Delta=0.22$~eV at $T\approx 50$~K) \cite{zzminaprl}.

Photoconduction (illumination-induced change of conduction due to nucleation of nonequilibrium electron-hole pairs) is a powerful method for energy structure study including detection of states whithin the gap. Here we apply this method for investigation of the Peierls gap in o-TaS$_3$. We demonstrate, that temperature- and electric-field-dependent energy levels inside the Peierls gap do really exist at low temperatures and are temperature and electric-field-dependent. We also argue that the low-temperature Peierls gap value in o-TaS$_3$ $2\Delta \gtrsim 0.2$~eV at $T \lesssim 50$~K.

\section{Experimental}
We have studied photoconduction spectra of nominally pure crystals provided by R.E. Thorne (Cornell University), and F.~Levy (Institute de Physique Appliqu\'ee, Lausanne). All the measuremens were performed in the two-contact configuration in the voltage-controlled regime. The sample quality was controlled by measuring the temperature variation of resistance and the threshold field value for onset of the nonlinear conduction. We studied thin crystals to enhance the relative contribution of photoconduction with respect to the bulk conduction. 4 samples (\# T1 --- T4) were picked up from the Thorne's batches and 6 samples (\# L1 --- L6) from the Levy' one. Samples' parameters are listed in Table~\ref{table}.

\begin{table*}
\begin{ruledtabular}
\caption{\label{table}Table 1. Samples' parameters}
\begin{tabular}{llllllll}
N of sample & Length, mm & $\sigma$, $\mu m^2$ & R$_{300 K}$, k$\Omega$ & Peaks energies, eV & $\hbar \omega ^*$ & Provided by: \\ 
L1 & 0.28 & 0.10 & 7.6 & 0.22 & 0.4 & Levy \\
L2 & 0.31 & 0.22 & 4.3 & 0.23 & 0.4 & Levy \\
L3 & 0.71 & 0.69 & 3.1 & 0.17, 0.18, 0.23&0.4& Levy \\
L4 & 0.50 & 0.22 & 7.0& 0.17, 0.18, 0.23 & 0.4& Levy \\
L5 & 0.16 & 0.06 & 7.6 & 0.20, 0.22& 0.4 & Levy \\
L6 & 0.40& 0.15 & 8.0 & 0.22& 0.4 & Levy \\
T1 \footnotemark[1]& 1.90 & 0.33 & 156.4 &0.20, 0.22 &0.44 & Thorne \\
T2 \footnotemark[1]& 0.39& 0.04 & 3.53& 0.15, 0.20 &0.44& Thorne \\
T3 & 0.4& 0.04& 26.4 &0.20, 0.22, 0.25 &0.44& Thorne \\
T4 & 1.2 &0.09 & 37.9&0.20, 0.22, 0.25 &0.41& Thorne \\
\end{tabular}
\end{ruledtabular}
\footnotetext[1]{Samples \# T1 and \# T2 were obtained by cleavage of the same crystal.}
\end{table*}

Photoconduction was measured using a grating monochromator with a globar as a light source. The light intensity was modulated at the low frequency (3.125~Hz and 6.25~Hz) by a light chopper, a lock-in amplifier was used for photocurrent measurements. To reduce the effect of light absorption by the air, the monochromator was evacuated to a pressure below 1~Torr. The air path between the output window of the monochromator and the input window of the cryostat was 30~cm. Therefore, the high-absorption regions around 0.29 and 0.36~eV were excluded from consideration. The samples were placed into a cryostat camera. The camera was filled by a heat-exchange gas (helium) at the normal pressure. A room-temperature window was made of KRS-5 and a low-temperature one was made of Si. All the spectra $S(\hbar\omega)$ presented here are normalized by the number of photons obtained from bolometer spectra $W(\hbar\omega)$ measured separately: $ S(\hbar\omega)\equiv \delta I(\hbar\omega) \hbar \omega/W(\hbar \omega)V$, where  $\delta I(\hbar\omega)$ is a photoconduction current, and $V$ is the voltage applied to a sample.

\section{Results}

\begin{figure} 
\epsfxsize=8.5cm
\leavevmode{\epsfbox{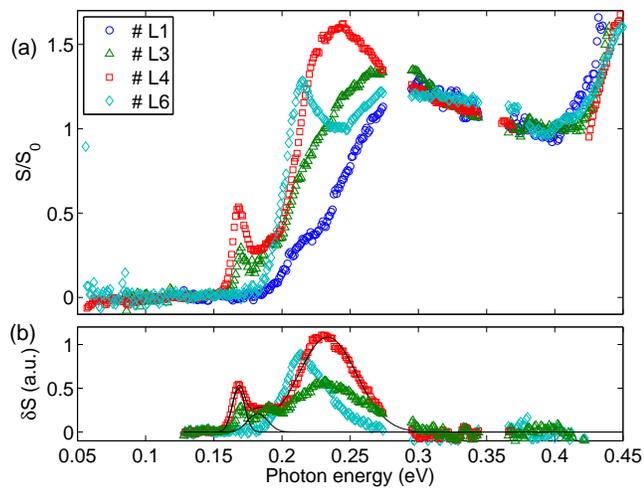}}
\caption{(a) Photoconduction spectra of four o-TaS$_3$ samples. For convenience all the spectra are normalized to unity at $\hbar \omega = 0.37$~eV; (b) subtraction spectra of photoconduction of three samples: \# L2, \# L3 and \# L4. Photoconduction spectrum of \# L1 sample is subtracted from others. Solid lines demonstrates a fit with a sum of gaussians and components of the sum.}
\label{spectra}
\end{figure}

Photoconduction at photon energies $\hbar \omega \gtrsim 0.2$~eV was found to be a nonmonotonous function of temperature and has a maximum at temperature $T_m \approx 50$~K, in agreement with earlier studies \cite{zzminaprl,zzminajetpl}. Because of the small value of the photoconduction current (typically few pA in maximum) we were able to obtain photoconduction spectra only in a restricted temperature range near the maximum. Linear conduction, which is the main source of the current noise, freezes out upon cooling. As a result, the temperature $T_{opt}$ at which the best signal/noise ratio is achieved is lower than $T_m$ and varies from sample to sample in the range $35{\rm~K}\lesssim T_{opt}\lesssim 45$~K. 

We found that decrease of the light intensity by a factor up to 8 (by means of grid filters) does not produce any change in the shape of photoconduction spectra. Similarly, no variation of the spectra shape due to doubling of the light chopper frequency was observed.

Fig.~\ref{spectra}(a) shows typical photoconduction spectra of the L-group samples. All the spectra were measured at voltages well below the threshold one for onset of CDW sliding. The results demonstrate a reasonably good spectra reproducibility at energies above 0.3 -- 0.35~eV. A growth of photoconduction is observed at photon energies above $\hbar \omega ^*\approx 0.4$~eV. A variety of structures is observed at energies below 300 meV. The most interesting features are peaks at energies $\hbar \omega < 300$~meV. Namely, we observed relatively narrow peaks  at energies 150, 170, 200 and 220~meV (HWHM~$\approx 5$~meV) and wider peaks at 230 -- 250~meV(HWHM~$\approx 20$~meV). Some samples exhibit weak features rather than peaks, at these energies, ({\em e.g.} the small bend at the 0.22 eV for the sample \#L1, Fig.\ref{spectra} (a)).  

In all the samples at least one of these peaks (or a weak feature at its position) may be distinguished in the photoconduction spectra. We suggested, therefore, that any spectrum is a superposition of spectral lines and a Peierls gap spectra. We also assume that there are no lines at energies 0.3 -- 0.4~eV. If so, then the difference between any couple of normalized spectra represents a set of lines.

Fig.~\ref{spectra}(b) shows typical results of subtraction of normalized spectra of o-TaS$_3$ samples. No features are observed at energies above 0.30 -- 0.35~eV. This indicates that the Peierls gap spectra does not contribute to the subtractive spectra. Peaks can be clearly distinguished at energies below 0.3~eV, the sum of 1-3 Gaussians fits any subtractive spectra quite reasonably.

Qualitatively similar behavior is also observed in the T-group samples. Quantitatively, there is a difference at least in the high-energy region: the photoconduction starts to grow at  $\hbar \omega^* \approx 0.45$~eV, instead of 0.4~eV as in the L-group samples (see Fig.~\ref{spectra}(a)). Similar differences, such as bigger values of peaks energies, than those for the L-group samples, and a shift of the edge of the photoconduction spectra to 0.22\,eV were observed also in the low-energy region. The origin of such differences is not clear yet.

\begin{figure} 
\epsfxsize=8.5cm
\leavevmode{\epsfbox{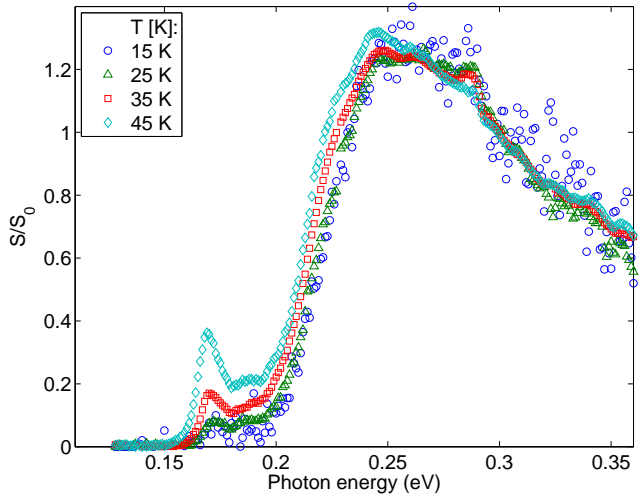}}
\caption{Typical temperature variation of photoconduction spectra of o-TaS$_3$ sample. For convenience all the spectra are normalized to unity at $\hbar \omega = 0.3$~eV. Sample \# L4.}
\label{spectra_vs_T}
\end{figure}

Fig.~\ref{spectra_vs_T} demonstrates a typical temperature variation of the spectrum shape. The high-energy part at $\hbar \omega > \hbar \omega_1 $ (where $\hbar \omega_1 = 0.3$~eV for the L-group samples and 0.35~eV for the T-group samples ) is practically independent of temperature within the accuracy of our measurements, whereas the low-energy one varies noticeably. The most significant evolution is a rapid decrease of the peaks amplitude with lowering temperature (see Fig.~\ref{spectra_vs_T}). Temperature dependences of the wide peaks are much weaker. Similar behavior is observed in all the samples.

\begin{figure} 
\epsfxsize=8.5cm
\leavevmode{\epsfbox{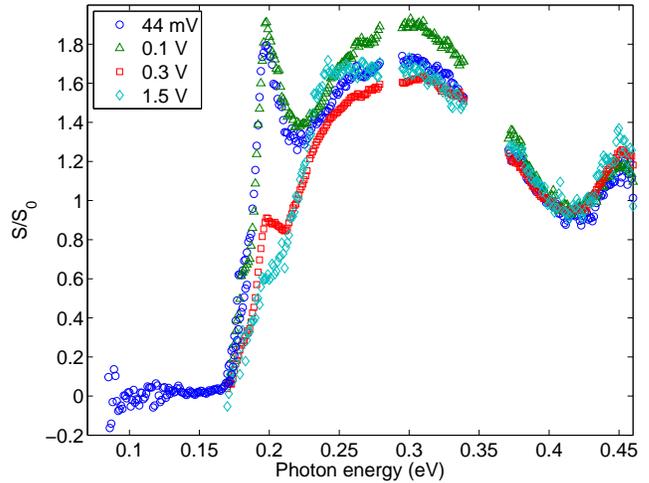}}
\caption{Photoconduction spectra of o-TaS$_3$ sample as a function of the voltage at $T = 30$~K. For convenience all the spectra are normalized to unity at $\hbar \omega = 0.37$~eV. Sample \#T4.}
\label{spectra_vs_V}
\end{figure}

One of the most interesting findings is dependence of the spectrum shape on the voltage applied to a crystal (Fig.~\ref{spectra_vs_V}). We observe this effect in the sample, which has an oustanding dependence of  photoconduction on the voltage: the photoconduction $\delta G$ increases up to one and a half order of magnitude with increasing the voltage from 100~mV (the boundary of the region of linear conduction) to 1~V (non-linear conduction region). One can see (see Fig.~\ref{spectra_vs_V}) that growth of the voltage from $V = 0.044$~V to $V = 0.3$~V leads to a noticeable  suppression of photoconduction at $\hbar \omega < 0.3$~eV and, in particular, to the suppression of the peak at $\hbar \omega = 0.2$~eV. This voltage range corresponds to onset of the smooth nonlinearity in the I-V curve. Further voltage ascending results in almost complete suppresion of this peak and simultaneous enhancement of photoconduction at energies  above 0.24 - 0.25~eV. No further spectrum variations were found for relatively large voltages above $V = 0.5$~V. Rapid growth of noise due to developing of slow motion of the CDW (CDW creep) complicates spectra study at larger fields. We do not observe any noticeable changes of the spectra at the high-energy part, at $E_1 > 0.35$~eV. The peak at 0.2~eV is suppressed with decreasing temperature, as well as the peaks in other samples.

\section{Discussion}

Observation of the energy states inside the Peierls gap is the main result of the present study. Similar maxima can be found in bolometric response spectra \cite{naditkis,ogap} at the energies 62, 100, 136 and 157 meV. The nature of this maxima was unclear except for the one at 62 meV, which was attributed to the amplitude soliton level in the middle of the Peierls gap (from \cite{naditkis,ogap} $2\Delta = 125$~eV was estimated). We would like to notice that any photon absorption process, including phonon excitation, gives a contribution to the bolometric response. In contrast, the study of photoconduction makes it possible to observe only electron processes. In the present study we do not find levels at these energies. An unusual feature of the observed levels is the electric field dependence of their intensity. The small value of the electric field ${\cal E}\ll E_i/l$, where $l$ is the free path length \cite{lcomment}, affecting energy levels proves participation of the collective state in the formation of the levels.

The energy states inside the Peierls gap may have various origins. First of all, these states may result from impurities and defects of the crystall lattice. A physical mechanism of formation of impurity levels in quasi-1D conductors is another than in semiconductors, where they are formed by orbital motion of electrons in the Coulomb potential of charged impurities. Here no orbital motion is possible and screening of charged impurities is strong and determined mostly by the collective mode. As it was noticed above, a charged impurity produces a couple of electronic levels inside the gap \cite{zavadowski}. Overlapping of this levels in samples with big amount of impurities can lead to reduction of effective value of the Peierls gap and probably to formation of long ``tails'' of states inside the gap \cite{brill,brillmint}.

The temperature variation of the amplitude of impurity lines (blanking out with lowering temperature) is similar to the phenomenon well-known for photoconduction in semiconductors. This blanking out means that impurity levels are localized. In this case photoexcitation of a current carrier into this localized state does not directly contribute to photoconduction. Further thermal excitation is required. As a probability of thermal excitation falls as $\exp(-E_i/kT)$, where $E_i$ is the depth of an impurity level (i.e. an energy distance to the edge of the Peierls gap), the peaks blank out with temperature decrease. The width of the peaks comply to the type of transition: the narrow peaks correspond to the level-level transitions ($\hbar \omega =2\Delta - 2E_i$, both initial and final states are localized), whereas the wide ones correspond to the level-gap transitions ($\hbar \omega =2\Delta - E_i$, either initial or final state is not localized). Consequently, the narrow peaks freeze-out, whereas the wide ones do not, in agreement with our observations. Though such a picture is in qualitative agreement with our studies, we cannot point out a couple of levels corresponding to such transitions. Beside this, the observed voltage dependence of energy levels is different. According to \cite{zavadowski} both amplitude and energy of levels created by a charged impurity depend on the CDW phase on the impurity cite and, therefore, are affected by the electric field. We observed only changing of amplitude of the levels whereas their energy position did not change.

The picture of the phenomenon may become more complicated if impurities form a superstructure \cite{coleman}. An additional modulation of the Peierls gap may appear in this case. Atoms of some impurities, for example, of adsorbed gases, may be able to diffuse at lengths of the order of atom spacing even at low temperatures. Then, if the CDW begin to move in the electric field, the position of those high-mobile impurities may change, that results to a change of the energy spectrum. Study of dedicated samples is required to establish a correspondence between the peaks and chemical impurities.

The CDW defects may also form energy levels inside the Peierls gap. Examples of such defects are amplitude and phase solitons, and dislocations of the CDW. Dislocations of the CDW may be a consequence of the big value of threshold field for onset of CDW sliding at low temperatures which is well-known for {\it o}-TaS$_3$ \cite{cdwreview}. One can expect that CDW motion  affects such states. Notice that in the present study we do not observe energy levels near the middle of the Peierls gap which we could ascribe to the amplitude solitons \cite{Braz}.

Recently two independent CDWs were found in {\it o}-TaS$_3$ at temperatures 50\,K --- 130\,K in the synchrotron X-ray study \cite{xray}. The existence of two or more CDWs at low temperatures in TaS$_3$ is enough to explain a part of the features in the spectra as a result of superposition of a few Peierls gaps edge spectra and also a set of levels corresponding to the edge dislocations arising between the CDWs with different values of the wave vector. But in this case one can expect a reasonable reproducibility of features in different samples which we do not observe (see Fig.~\ref{spectra}).

In conclusion, we demonstrate, by using photoconduction, that temperature- and electric-field-dependent energy levels inside the Peierls gap do really exist. We also show that the low-temperature Peierls gap value, $2\Delta $, in o-TaS$_3$ is $\gtrsim 0.2$~eV. Further studies are required to clarify the nature of the spectral features found.

We are grateful to S.~N.~Artemenko and V.~Ya.~Pokrovskii for useful discussions, R.~E.~Thorne and F.~Levy for providing high-quality crystals and V.~E.~Minakova for help in sample preparations. The work was supported by RFBR (grant \#07-02-01131, \#06-02-72552). These researches were performed in the frame of the CNRS-RAS-RFBR Associated European Laboratory ``Physical properties of coherent electronic states in condensed matter'' between Institut N\'eel, CNRS and IRE RAS.


\begin{thebibliography}{99}
\bibitem{cdwreview}P. Monceau, in: ``Electronic Properties of Inorganic Quasi-one-dimensional Conductors'', Part 2. Ed. by P. Monceau. Dortrecht: D. Reidel Publ. Comp., 1985; G. Gr\"{u}ner, Rev. Mod. Phys. {\bf 60}, 1129 (1988).

\bibitem{zavadowski}I. T\"{u}tt\"{o} and A. Zawadowski, Phys. Rev. B {\bf 32}, 2449 (1985).

\bibitem{sambongi}T. Sambongi, K. Tsutsumi, Y. Shiozaki, M. Yamamoto, K. Yamaya, and Y. Abe, Solid State Commun. {\bf 22}, 729 (1977).

\bibitem{TaS3hall}Yu.I.~Latyshev, Ya.S.~Savitskaya, V.V.~Frolov, Pis'ma ZhETF, 
{\bf 38}, 446 (1983).

\bibitem{naditkis}F.Ya. Nad, M.E.Itkis,   Pis'ma v ZhETP, {\bf 63}, 246, (1996).

\bibitem{brill} S. L. Herr, G. Minton, and J.W. Brill, Phys. Rev. B {\bf 33}, 8851 (1986).

\bibitem{latnewgap}Yu.I. Latyshev, P. Monceau, S. Brazovskii, A.P. Orlov, and T. Fournier, Phys. Rev. Lett. {\bf 96}, 116402 (2006).

\bibitem{zzminaprl}S.V. Zaitsev-Zotov, V.E. Minakova, Phys. Rev. Lett. {\bf97}, 266404 (2006).

\bibitem{zzminajetpl}S.V.~Zaitsev-Zotov, V.E.~Minakova, Pis'ma v ZhETP, 
{\bf 79}, 680 (2004) [JETP Letters, {\bf 79}, 550 (2004)].

\bibitem{ogap}M.E.~Itkis, F.Ya.~Nad,  Pis'ma v ZhETP, {\bf 39}, 373 (1984).

\bibitem{lcomment} For $l \lesssim 1$\,$\mu$m $E_i/l \gtrsim 10^3$\,V/cm.

\bibitem{brillmint} G. Minton, and J.W. Brill, Sol. St. Commun.,
{\bf65}, 1069 (1988).

\bibitem{coleman}Q. Xue, Z. Dai, Y. Gong, C. G. Slough and R. V. Coleman, Phys. Rev. B {\bf 48}, 1986 (1993).

\bibitem{Braz} S. Brazovski, JETP Letters, {\bf 78}, 677 (1980).

\bibitem{xray} I. Katsuhiko, T. Masakatsu, H. Kazuki, I. Koichi, T. Satoshi, Y. Kenichiro, H. Noriaki, I. Naoshi, N. Yoshio, I. Takayoshi, T. Hidenori, Journal of the Physical Society of Japan, {\bf 77}, 093708 (2008).


\end{thebibliography}
\end{document}